\documentstyle[11pt,dina4p,twoside]{article}
\begin{document}
\parskip5pt plus2pt minus2pt

\title{\rightline{\normalsize DESY 94-173}\vskip 1cm
       Decomposition of Representations of
CAR Induced by Bogoliubov Endomorphisms}
\author{Jens B{\"o}ckenhauer\\II. Institut f{\"u}r
Theoretische Physik, Universit{\"a}t Hamburg\\
Luruper Chaussee 149, D-22761 Hamburg}
\maketitle

\begin{abstract}
In a Fock representation, a non-surjective
Bogoliubov transformation of CAR leads to
a reducible representation. For the case that
the corresponding Bogoliubov operator has
finite corank, the decomposition into
irreducible subrepresentations is
clarified. In particular, it turns out that
the number of appearing subrepresentations
is completely determined by the corank.
\end{abstract}

\newtheorem{definition}{Definition}[section]
\newtheorem{lemma}[definition]{Lemma}
\newtheorem{corollary}[definition]{Corollary}
\newtheorem{theorem}[definition]{Theorem}

\section{Introduction}
Unitary i.e.~surjective Bogoliubov operators $U$
correspond to Bogoliubov automorphisms $\varrho_U$
of the canonical anticommutation relations
(CAR). In a Fock representation $\pi_P$,
Bogoliubov automorphisms lead to irreducible
representations. More precisely, the representation
$\pi_P\circ\varrho_U$ is again a Fock representation.
The unitary equivalence class of $\pi_P\circ\varrho_U$
is in a certain way classifiable by the operator
$U^*PU$ \cite{Ara1,Ara2,PS}.

Here we study Bogoliubov endomorphisms $\varrho_V$,
where $V$ is a non-surjective Bogoliubov operator.
Throughout this paper, we consider a separable,
infinite dimensional Hilbert space ${\cal K}$ with
complex conjugation $\Gamma$, and we use Araki's
formalism of the selfdual CAR algebra
${\cal C}({\cal K},\Gamma)$ which is equivalent to
the more familiar notion of Clifford algebras over
real Hilbert spaces \cite{Ara2,Bin}.
We suppose that the corank of
the Bogoliubov operator is finite. Since $V$ is
non-surjective, a representation of the form
$\pi_P\circ\varrho_V$ is reducible. It turns out
that the decomposition into irreducibles is
determined by the corank of $V$. If it is an
even number, say $2N$, we prove that
$\pi_P\circ\varrho_V$ decomposes into $2^N$
mutually equivalent Fock representations.
The explicit form of those subrepresentations
can be seen from our proofs. On the other hand,
if the corank of $V$ is an odd number, say
$2N+1$, we find that $\pi_P\circ\varrho_V$
decomposes into $2^{N+1}$ irreducibles where
we have two equivalence classes of $2^N$
mutually equivalent subrepresentations each.
Furthermore, we investigate what happens when
those representations become restricted to the
even subalgebra ${\cal C}({\cal K},\Gamma)^+$
of ${\cal C}({\cal K},\Gamma)$, in both cases.
It turns out that the situation then becomes
inverted somehow.

\section{Preliminaries}
As the underlying test function space, we consider
a separable, infinite dimensional Hilbert space
${\cal K}$ with an antiunitary involution $\Gamma$
(complex conjugation), $\Gamma^2={\bf 1}$, fulfilling
\[ \langle \Gamma f, \Gamma g \rangle = \langle g ,
f \rangle, \qquad f,g\in{\cal K}. \]
Let ${\cal C}_0({\cal K},\Gamma)$ be the unital
$\ast$-algebra which is algebraically generated
by the range of the linear embedding
$B:{\cal K} \rightarrow {\cal C}_0({\cal K},\Gamma)$,
where the following relations hold,
\[ B(f)^*=B(\Gamma f),\qquad
\{ B(f)^*,B(g) \} = \langle f,g \rangle {\bf 1}. \]
There is a unique $C^*$-norm of
${\cal C}_0({\cal K},\Gamma)$ which satisfies
\[ \|B(f)\|=\frac{1}{\sqrt{2}}\sqrt{\|f\|^2
+\sqrt{\|f\|^4-|\langle f,\Gamma f\rangle|^2}}.\]
The $C^*$-completion is referred as the selfdual
CAR algebra over $({\cal K},\Gamma)$ and is denoted
by ${\cal C}({\cal K},\Gamma)$. Elements of the
set
\[ {\cal I}({\cal K},\Gamma) = \{ V\in {\cal B}
({\cal K}) \,|\, [V,\Gamma]=0,\,\, V^*V={\bf 1} \} \]
of $\Gamma$ commuting isometries on ${\cal K}$ are
called Bogoliubov operators. Each Bogoliubov
operator $V\in {\cal I}({\cal K},\Gamma)$ induces a
unital $\ast$-endomorphism $\varrho_V$ of
${\cal C}({\cal K},\Gamma)$, defined by its action
on generators,
\[ \varrho_V(B(f))=B(Vf), \]
$\varrho_V$ is precisely an automorphism if $V$ is
surjective, i.e. ker$V^*=\{0\}$. In this paper, we
consider Bogoliubov operators with dim ker$V^*
<\infty$ which are Fredholm operators; we set
\[ M_V= \mbox{dim ker}V^*= \mbox{dim coker}V=
\mbox{index}V^* = -\mbox{index}V \]
and define subsets of ${\cal I}({\cal K},\Gamma)$,
\[ {\cal I}^n({\cal K},\Gamma)= \{ V \in
{\cal I}({\cal K},\Gamma)\,\,|\,\,M_V=n \}. \]
Now we come to the states of
${\cal C}({\cal K},\Gamma)$ we are interested in.
\begin{definition}
\label{quasifree}
A state $\omega$ of ${\cal C}({\cal K},\Gamma)$ is called
quasifree, if for all $n\in {\bf N}$
\begin{eqnarray}
\omega (B(f_1) \cdots B(f_{2n+1})) &=& 0, \\
\label{perm1}
\omega (B(f_1) \cdots B(f_{2n})) &=&
(-1)^\frac{n(n-1)}{2} \sum_\sigma
\mbox{\rm sign} \sigma\prod_{j=1}^n \omega
(B(f_{\sigma(j)})B(f_{\sigma(n+j)}))
\end{eqnarray}
holds. The sum runs over all permutations
$\sigma \in {\cal S}_{2n}$ with the property
\begin{equation}
\label{perm2}
\sigma (1) < \sigma (2) < \cdots < \sigma (n),
\qquad \sigma (j) < \sigma (j+n), \qquad j=1,\ldots,n.
\end{equation}
\end{definition}
Quasifree states are therefore completely characterized by
their two point functions. Moreover, it is known that there
is a one to one correspondence between the set of quasifree
states and the set
\[ Q({\cal K},\Gamma)=\{S\in{\cal B}({\cal K})\,|\,
S=S^*,\, 0 \le S \le {\bf 1},\,
S + \Gamma S \Gamma = {\bf 1} \}, \]
given by the formula
\begin{equation}
\label{phi-S}
\omega (B(f)^* B(g)) = \langle f,Sg \rangle .
\end{equation}
The quasifree state characterized by Eq.~(\ref{phi-S})
will be referred as $\omega_S$. A quasifree state,
composed with a Bogoliubov endomorphism is again a
quasifree state, namely we have
$\omega_S\circ\varrho_V=\omega_{V^*SV}$.
The projections in $Q({\cal K},\Gamma)$
are called basis projections. For a basis projection $P$,
the state $\omega_P$ is pure and is called a
Fock state. The corresponding GNS representation
$({\cal H}_P,\pi_P,|\Omega_P\rangle)$
is irreducible, it is called the Fock representation.
The vector $|\Omega_P\rangle\in{\cal H}_P$ is
called the Fock vacuum and the representation space
${\cal H}_P$ is precisely the antisymmetric Fock space
over $P{\cal K}$. Araki proved \cite{Ara1} the following
\begin{lemma}
\label{Fock}
Let $P$ be a basis projection and let $\omega$ be a
state of ${\cal C}({\cal K},\Gamma)$ which satisfies
\begin{equation}
\omega(B(f)B(f)^*)=0, \qquad f\in P{\cal K}.
\end{equation}
Then $\omega$ is the Fock state $\omega=\omega_P$.
\end{lemma}
Powers and St{\o}rmer \cite{PS} developed an important
criterion for the quasiequivalence of representations being
induced by gauge invariant quasifree states. Araki \cite{Ara1}
generalized this criterion for arbitrary quasifree states.
\begin{theorem}
\label{Hilbert Schmidt}
Two quasifree states $\omega_{S_1}$ and
$\omega_{S_2}$ of ${\cal C}({\cal K},\Gamma)$
induce quasiequivalent representations, if and only if
the difference $S_1^\frac{1}{2} - S_2^\frac{1}{2}$ is
Hilbert Schmidt class.
\end{theorem}
In this paper, we study representations of the form
$\pi_P\circ\varrho_V$. If $V$ is surjective, i.e.
$V\in{\cal I}^0({\cal K},\Gamma)$, then $\varrho_V$
is an automorphism, $V^*PV$ again a basis projection
and $\pi_P\circ\varrho_V$ again an (irreducible) Fock
representation, $\pi_P\circ\varrho_V=\pi_{V^*PV}$.
But if $V\in{\cal I}^n({\cal K},\Gamma)$, $n>0$, then
we will see that $\pi_P\circ\varrho_V$ is always
reducible. We are interested in the decomposition
into its irreducible subrepresentations and their
structure. The first step in this investigation is
the following Lemma \ref{Binnenhei} which, in the
frame of common CAR formalism, was proven by
Rideau \cite{Rideau}, and, in the frame of the
selfdual CAR formalism, has been formulated by
Binnenhei \cite{Bin,Binnenneu}. We define
\[ N_V=\mbox{dim}(\mbox{ker}V^* \cap P{\cal K}), \]
$0\le N_V \le \frac{1}{2} M_V$, and we choose an
orthonormal basis (ONB) $\{ k_n,n=1,2,\ldots,N_V \}$
of the space ker$V^*\cap P{\cal K}$. Further we
introduce the set $I_{N_V}$ of multi-indices
\[ I_{N_V} = \{ \beta=(\beta_1,\beta_2,\ldots,
\beta_l) \,|\, 1 \le l \le N_V,\,\,
\beta_1,\ldots,\beta_l \in {\bf N},\,\,
1\le\beta_1<\beta_2<\cdots<
\beta_l\le N_V \} \cup \{ 0 \}. \]
We remark that $I_{N_V}$ consists of $2^{N_V}$
elements. Next we set
$a_j = T_P(-1) \pi_P (B(k_j))$,
$j=1,2,\ldots,N_V$, where
$T_P(-1)$ denotes the unitary operator in
${\cal B}({\cal H}_P)$, unique up to a phase,
which implements the automorphism $\alpha_{-1}$
of ${\cal C}({\cal K},\Gamma)$, defined by its
action on generators,
\begin{equation}
\label{alpha}
\alpha_{-1}(B(f))=-B(f)
\end{equation}
in the Fock representation $\pi_P$. The implementing
$T_P(-1)$ exists because $\alpha_{-1}$ leaves any
quasifree state invariant \cite{Ara1}.
We define for $\beta\in I_{N_V}$
\[ A_\beta = a_{\beta_1} a_{\beta_2} \cdots a_{\beta_l},
\qquad \beta\neq 0,\qquad A_0={\bf 1};\qquad
|\Omega_\beta\rangle = A_\beta|\Omega_P\rangle. \]
By ${\cal H}_\beta$ we denote the subspaces of
${\cal H}_P$ being generated by the action of
$\pi_P\circ\varrho_V({\cal C}({\cal K},\Gamma))$
on vectors $|\Omega_\beta\rangle$, and by
$\pi_{V^*PV}^{(\beta)}$ the restrictions of
$\pi_P\circ\varrho_V$ to ${\cal H}_\beta$,
$\beta\in I_{N_V}$.
\begin{lemma}
\label{Binnenhei}
Let $\omega_P$ be a Fock state,
$({\cal H}_P,\pi_P,|\Omega_P\rangle)$ the
corresponding Fock representation and
$\varrho_V$ a Bogoliubov endomorphism of
${\cal C}({\cal K},\Gamma)$,
$V\in{\cal I}({\cal K},\Gamma)$. Then the
representation $\pi_P\circ\varrho_V$ decomposes
into $2^{N_V}$ cyclic subrepresentations,
\begin{equation}
\pi_P\circ\varrho_V = \bigoplus_{\beta\in I_{N_V}}
\pi_{V^*PV}^{(\beta)}
\end{equation}
where $({\cal H}_\beta,\pi_{V^*PV}^{(\beta)},
|\Omega_\beta\rangle)$ are GNS representations of
the state $\omega_P\circ\varrho_V=\omega_{V^*PV}$.
\end{lemma}
However, this decomposition is a decomposition into
cyclic representations, but in general not into
irreducibles because $V^*PV$ is in general not a
basis projection. In our analysis of the decomposition
of representations $\pi_P\circ\varrho_V$, the number
$M_V$ turns out to be the important quantity.
It is useful to distinguish the even case, $M_V=2N$,
and the odd case, $M_V=2N+1$.

\section{The Even Case: $M_V=2N$}
Let us begin our investigation with the special case
that $V\in{\cal I}^2({\cal K},\Gamma)$, i.e.
$M_V=2$. Then we choose an ONB
$\{e_+,e_-\}$ of ker$V^*$ with the property
$e_+=\Gamma e_-$. This is always possible because
ker$V^*$ is $\Gamma$-invariant since $[V,\Gamma]=0$.
Following Araki \cite{Ara1,Ara2}, in a
$\Gamma$-invariant space it is always possible to
choose a $\Gamma$-invariant ONB, say here $f_1$ and
$f_2$, $\Gamma f_j=f_j$, $j=1,2$. Then vectors
\[ e_\pm = \frac{1}{\sqrt{2}}(f_1 \pm if_2) \]
have the required property. Now we have a look at
the numbers $\lambda_+,\lambda_-\in [0,1]$,
\[ \lambda_\pm = \langle e_\pm,Pe_\pm \rangle. \]
Since $P+\Gamma P\Gamma ={\bf 1}$ we have
$\lambda_++\lambda_-=1$,
\[ \lambda_++\lambda_-=\langle e_+,Pe_+\rangle +
\langle e_-,Pe_-\rangle = \langle e_+, Pe_+ \rangle +
\langle \Gamma e_+,P\Gamma e_+ \rangle =
\langle (P+\Gamma P\Gamma) e_+,e_+ \rangle =1. \]
Suppose first that one of these numbers equals
zero, say $\lambda_-=0$, and therefore
$\lambda_+=1$. We then have the case
$e_+\in P{\cal K}$ and $e_-\in ({\bf 1}-P)
{\cal K}$. The state $\omega_P\circ\varrho_V=
\omega_{V^*PV}$ is then pure (a Fock state) because
$V^*PV$ is a basis projection; namely we find
\begin{eqnarray*}
(V^*PV)^2 &=& V^*PVV^*PV = V^*P({\bf 1}-|e_+\rangle
\langle e_+|-|e_-\rangle\langle e_-|)PV \\
&=& V^*PV - V^*|e_+\rangle\langle e_+|V
=V^*PV.
\end{eqnarray*}
But by Lemma \ref{Binnenhei} it follows that the
representation $\pi_P\circ\varrho_V$ decomposes into
two equivalent irreducible (Fock) representations,
\[ \pi_P\circ\varrho_V=\pi_{V^*PV}^{(1)} \oplus
\pi_{V^*PV}^{(2)}. \]
The corresponding representation spaces
${\cal H}_{V^*PV}^{(1)}$ and ${\cal H}_{V^*PV}^{(2)}$
are generated by the action of
$\pi_P\circ\varrho_V({\cal C}({\cal K},\Gamma))$
on vectors $|\Omega^{(1)}\rangle$ and
$|\Omega^{(2)}\rangle$, respectively,
\begin{equation}
\label{globvec}
|\Omega^{(1)}\rangle = |\Omega_P\rangle, \qquad
|\Omega^{(2)}\rangle = a_+ |\Omega_P\rangle,
\qquad a_+ = T_P(-1)\pi_P(B(e_+)).
\end{equation}
Now suppose the case that both numbers
$\lambda_\pm\neq 0$ and
therefore $\lambda_\pm\neq 1$. We argue that in this
case $N_V=0$, i.e. ker$V^*\cap P{\cal K}=\{0\}$:
Suppose there is a $v\in$ker$V^*\cap P{\cal K}$.
Since $v\in$ker$V^*$ we have to write
\[ v=\mu_+ e_+ + \mu_- e_-, \]
with complex numbers $\mu_\pm \in {\bf C}$. On
the other hand, since $v\in P{\cal K}$ it follows
$\Gamma v\in ({\bf 1}-P){\cal K}$ and therefore
\[ \langle \Gamma v, P \Gamma v \rangle = 0. \]
We have
$\Gamma v= \bar{\mu}_+ e_- + \bar{\mu}_-e_+$,
and since $\langle e_\pm,P e_\mp \rangle =0$ by
\[ \langle e_+,P e_- \rangle = \langle e_+,
({\bf 1}-\Gamma P \Gamma) e_- \rangle = - \langle
e_+,\Gamma P e_+ \rangle = - \langle Pe_+, \Gamma
e_+ \rangle = - \langle e_+, P e_- \rangle \]
this reads
\[ \langle \Gamma v,P\Gamma v \rangle =
|\mu_+|^2 \langle e_-,P e_- \rangle +
|\mu_-|^2 \langle e_+, P e_+ \rangle =
|\mu_+|^2 \lambda_- + |\mu_-|^2 \lambda_+=0. \]
But since $\lambda_\pm$ both are non-zero and
positive by assumption this establishes
$\mu_+=\mu_-=0$ and therefore $v=0$.
By Lemma \ref{Binnenhei} we obtain that in this case
$({\cal H}_P,\pi_P\circ\varrho_V,|\Omega_P\rangle)$
is a GNS representation of the
state $\omega_P\circ\varrho_V$. However, as we will
see, in this case the state is a mixture \cite{ich}.
In the following we use the fact that if a state
$\omega$ is a mixture of two pure states
$\omega_1$ and $\omega_2$ of a $C^*$-algebra,
$\omega_1\neq\omega_2$,
\[ \omega = \lambda_1 \omega_1 + \lambda_2 \omega_2,
\qquad \lambda_1,\lambda_2>0,\qquad \lambda_1 +
\lambda_2 =1, \]
with associated GNS representations
$({\cal H}_j,\pi_j,|\Omega_j\rangle)$, $j=1,2$,
a GNS representation for $\omega$ is given by
the direct sum
\[ ({\cal H}_1 \oplus {\cal H}_2,\pi_1 \oplus
\pi_2,\sqrt{\lambda_1}|\Omega_1\rangle \oplus
\sqrt{\lambda_2}|\Omega_2\rangle ). \]
\begin{lemma}
\label{Bockenhauer}
Let $\omega_P$ be a Fock state and let
$({\cal H}_P,\pi_P,|\Omega_P\rangle)$ be the corresponding
Fock representation of ${\cal C}({\cal K},\Gamma)$. Let
$V\in {\cal I}^2({\cal K},\Gamma)$ be a Bogoliubov operator
and choose an ONB $\{e_+,e_-\}$ of {\rm ker}$V^*$ with the
property that $e_+=\Gamma e_-$. If both numbers
\begin{equation}
\lambda_\pm = \langle e_\pm, P e_\pm \rangle \neq 0,
\end{equation}
then the state $\omega_P\circ\varrho_V$ is a mixture of
two Fock states $\omega_{P_\pm}$,
\begin{equation}
\label{decom}
\omega_P\circ\varrho_V=\lambda_+\omega_{P_+} +
\lambda_-\omega_{P_-}
\end{equation}
where the basis projections $P_+$ and $P_-$ are
explicitely given by
\begin{equation}
\label{basis}
P_\pm=V^*PV+\lambda_\pm^{-1}V^*P(E_\mp-E_\pm)PV,
\qquad E_\pm=|e_\pm\rangle\langle e_\pm|.
\end{equation}
Moreover, the representation $\pi_P\circ\varrho_V$ is cyclic
and decomposes therefore into a direct sum of two
irreducible representations which are equivalent.
\end{lemma}
{\it Proof.} Since the orthonormal vectors
$e_+,e_-$ span the kernel
of $V^*$ (i.e. the cokernel of $V$),
generators $B(e_+)$ and $B(e_-)$ anticommute
with each generator $B(Vf)$, $f\in{\cal K}$.
Therefore operators
\[ a_\pm = T_P(-1) \pi_P(B(e_\pm)) \]
lie in the commutant of
$\pi_P\circ\varrho_V({\cal C}({\cal K},\Gamma))$,
\[ a_\pm \in \pi_P\circ\varrho_V
({\cal C}({\cal K},\Gamma))'. \]
Since $\lambda_\pm \neq 0$ we have the well defined,
normed vectors in ${\cal H}_P$,
\begin{equation}
\label{locvec}
|\Omega_\pm\rangle = \lambda_\pm^{-\frac{1}{2}} a_\pm
|\Omega_P \rangle.
\end{equation}
We define states $\omega_+,\omega_-$ of
${\cal C}({\cal K},\Gamma)$ by
\[ \omega_\pm(x)=\langle \Omega_\pm
|\pi_P\circ\varrho_V(x)
|\Omega_\pm\rangle = \lambda_\pm^{-1}
\langle \Omega_P| \pi_P\circ\varrho_V(x)
\pi_P(B(e_\mp)B(e_\pm)) |\Omega_P \rangle,
\qquad x\in {\cal C}({\cal K},\Gamma),\]
such that we find
\[ \omega_P\circ\varrho_V=\lambda_+\omega_+
+\lambda_-\omega_- \]
by $\{B(e_+),B(e_-)\}={\bf 1}$. We are able to compute the
two point functions of $\omega_+$ and $\omega_-$ by
reading the permutation formulae (\ref{perm1}), (\ref{perm2})
for the quasifree state $\omega_P$,
\begin{eqnarray*}
\omega_\pm (B(f)B(g)) &=&
\lambda_\pm^{-1} \langle \Omega_P|\pi_P\circ\varrho_V
(B(f)B(g))\pi_P(B(e_\mp)B(e_\pm))
|\Omega_P \rangle \\
&=& \lambda_\pm^{-1} \omega_P(B(Vf)B(Vg)
B(e_\mp)B(e_\pm)) \\
&=& \lambda_\pm^{-1} \omega_P(B(Vf)B(Vg))
\omega_P(B(e_\mp)B(e_\pm)) \\
&& \qquad +
\lambda_\pm^{-1} \omega_P(B(Vf)B(e_\pm))
\omega_P(B(Vg)B(e_\mp)) \\
&& \qquad -
\lambda_\pm^{-1} \omega_P(B(Vf)B(e_\mp))
\omega_P(B(Vg)B(e_\pm)) \\
&=& \langle \Gamma f, V^*PVg \rangle +
\lambda_\pm^{-1} \langle \Gamma f,V^*Pe_\pm \rangle
\langle \Gamma g, V^*Pe_\mp \rangle \\
&& \qquad -
\lambda_\pm^{-1} \langle \Gamma f,V^*Pe_\mp \rangle
\langle \Gamma g, V^*Pe_\pm \rangle.
\end{eqnarray*}
Since $V^*e_\pm=0$ and $[V,\Gamma]=0$ we find
\[ \langle \Gamma g, V^*Pe_\pm \rangle = \langle
V^*\Gamma Pe_\pm,g \rangle = \langle V^*(\Gamma-
P\Gamma) e_\pm,g \rangle = - \langle V^*Pe_\mp,g
\rangle = -\langle e_\mp, PVg \rangle. \]
Hence we can write
\[ \omega_\pm(B(f)B(g))= \langle \Gamma f,
P_\pm g \rangle, \]
where
\[ P_\pm=V^*PV+\lambda_\pm^{-1}V^*P(E_\mp-E_\pm)PV,
\qquad E_\pm=|e_\pm\rangle\langle e_\pm|. \]
Using $V^*E_\pm=0=E_\pm V$ and $\Gamma E_\pm=
E_\mp \Gamma$, one finds easily the relation
\[ P_\pm + \Gamma P_\pm \Gamma ={\bf 1}, \]
namely we compute
\begin{eqnarray*}
P_\pm + \Gamma P_\pm \Gamma &=& V^*PV +
\lambda_\pm^{-1} V^*P(E_\mp-E_\pm)PV + V^*({\bf 1}
-P)V \\ && \qquad \qquad
+ \lambda_\pm^{-1}V^*({\bf 1}-P)\Gamma(E_\mp
-E_\pm)\Gamma({\bf 1}-P)V \\
&=& {\bf 1} + \lambda_\pm^{-1} V^*P(E_\mp-E_\pm)PV
+ \lambda_\pm^{-1} V^*P(E_\pm-E_\mp)PV = {\bf 1}.
\end{eqnarray*}
In the next step we show that $P_\pm^2=P_\pm$ i.e. that
$P_+$ and $P_-$ are basis projections. For simplicity
we check at first only the case $P_+^2=P_+$.
We begin with some helpful formulae. Since
$E_++E_-$ is the projection onto the kernel of
$V^*$ we have
\begin{equation}
\label{kern}
VV^*={\bf 1}-E_--E_+.
\end{equation}
By $\lambda_\pm=\langle e_\pm,Pe_\pm \rangle$ we get
\begin{equation}
\label{sigma}
E_\pm PE_\pm = \lambda_\pm E_\pm,
\end{equation}
and since $\langle e_\pm,P e_\mp \rangle =0$ we find
\begin{equation}
\label{null}
E_+PE_-=E_-PE_+=0.
\end{equation}
Define
\[ P_{+,1}=V^*PV,\qquad P_{+,2}=-\lambda_+^{-1}
V^*PE_+PV, \qquad P_{+,3}=\lambda_+^{-1}V^*PE_-PV \]
such that
\[ P_+=P_{+,1}+P_{+,2}+P_{+,3}. \]
We obtain the following list of products by using
Eqs.~(\ref{kern}), (\ref{sigma}) and (\ref{null}).
\begin{eqnarray*}
P_{+,1}^2 &=& V^*PV-V^*PE_+PV-V^*PE_-PV, \\
P_{+,1}P_{+,2} &=& (1-\lambda_+^{-1})V^*PE_+PV,\\
P_{+,1}P_{+,3} &=& \lambda_+^{-1}(1-\lambda_-)V^*PE_-PV,\\
P_{+,2}P_{+,1} &=& (1-\lambda_+^{-1})V^*PE_+PV, \\
P_{+,2}^2 &=& (\lambda_+^{-1}-1)V^*PE_+PV, \\
P_{+,2}P_{+,3} &=& 0, \\
P_{+,3}P_{+,1} &=& \lambda_+^{-1}(1-\lambda_-)V^*PE_-PV,\\
P_{+,3}P_{+,2} &=& 0, \\
P_{+,3}^2 &=& \lambda_+^{-2}\lambda_-(1-\lambda_-)V^*PE_-PV.
\end{eqnarray*}
By using only $\lambda_++\lambda_-=1$ we compute
\begin{eqnarray*}
P_+^2 &=& \sum_{k,l=1}^3 P_{+,k}P_{+,l} \\
&=& V^*PV + (-1+1- \lambda_+^{-1}+1 - \lambda_+^{-1}
+ \lambda_+^{-1} -1) V^*PE_+PV \\
&& \qquad + (-1+1+1+\lambda_+^{-1}\lambda_-) V^*PE_-PV \\
&=& V^*PV - \lambda_+^{-1}V^*PE_+PV + \lambda_+^{-1}
V^*PE_-PV \\
&=& P_+.
\end{eqnarray*}
By interchanging all $+$ and $-$ indices this reads
$P_-^2=P_-$. We have proven that $P_+$ and $P_-$ are
both basis projections, hence the states $\omega_+$ and
$\omega_-$ satisfy
\[ \omega_\pm (B(f)B(f)^*)=0, \qquad
f\in P_\pm {\cal K}, \]
and therefore they are Fock states $\omega_\pm=
\omega_{P_\pm}$ by Lemma \ref{Fock}. As already
mentioned,
$({\cal H}_P,\pi_P\circ\varrho_V,|\Omega_P\rangle)$
is a GNS representation of the state
$\omega_P\circ\varrho_V$ by Lemma \ref{Binnenhei}.
According to the decomposition (\ref{decom}) of this
state, its GNS representation therefore splits into two
Fock representations. Finally we emphasize that these
Fock representations are equivalent: The difference
\[ P_+-P_-=\lambda_+^{-1}\lambda_-^{-1}V^*P
(E_--E_+)PV \]
is obviously Hilbert Schmidt class because $E_-$
and $E_+$ are rank-one-projections. Using Theorem
\ref{Hilbert Schmidt} one finds that $\omega_{P_+}$
and $\omega_{P_-}$ give rise to equivalent
representations, q.e.d.

We observe the somewhat amazing phenomenon that in
the case that one of the numbers $\lambda_+$ and
$\lambda_-$ vanishes, the state
$\omega_P\circ\varrho_V$ remains pure but the
representation
$({\cal H}_P,\pi_P\circ\varrho_V,|\Omega_P\rangle)$
is no longer cyclic; it splits into two equivalent
irreducibles, and, on the other hand, if
$\lambda_\pm\neq 0$ both, then the representation
$({\cal H}_P,\pi_P\circ\varrho_V,|\Omega_P\rangle)$
remains cyclic but the state
$\omega_P\circ\varrho_V$ becomes a mixture of two
pure states. In both cases we find that
$\pi_P\circ\varrho_V$ decomposes into
two equivalent irreducible (Fock) representations;
this fact is true for each basis projection $P$
and each $V\in{\cal I}^2({\cal K},\Gamma)$ and will
be used for proving the main result of this section.
\begin{theorem}
\label{even}
Let $\omega_P$ be a Fock state and let $({\cal H}_P,
\pi_P,|\Omega_P\rangle)$ be the corresponding Fock
representation of the selfdual CAR algebra
${\cal C}({\cal K},\Gamma)$.
Further let $V\in{\cal I}({\cal K},\Gamma)$ be a
Bogoliubov operator with finite even corank, i.e.
$M_V=2N$, $N\in{\bf N}_0$, and let $\varrho_V$
be the corresponding Bogoliubov
endomorphism. Then the representation
$\pi_P\circ\varrho_V$ decomposes into $2^N$ mutually
equivalent irreducible (Fock) representations.
\end{theorem}
{\it Proof.} We choose a $\Gamma$-invariant ONB
$\{e_n,n\in{\bf N}\}$ of ${\cal K}$, i.e.
$e_n=\Gamma e_n$, $n\in{\bf N}$. Further we choose
a $\Gamma$-invariant ONB $\{f_n,n=1,2,\ldots,2N\}$ of ker$V^*$,
$f_n=\Gamma f_n$, $n=1,2,\ldots,2N$.
Moreover, we define
\[ f_{2N+n}= Ve_n, \qquad n\in {\bf N}. \]
Since ${\cal K}$ is the direct sum of the range of $V$
and the kernel of $V^*$, the set $\{f_n,n\in{\bf N}\}$
forms another $\Gamma$-invariant ONB of ${\cal K}$ and
we can write
\[ V=\sum_{n=1}^\infty |f_{2N+n}\rangle\langle e_n|. \]
We define Bogoliubov operators, the unitary
$V_0\in{\cal I}^0({\cal K},\Gamma)$, and
$V_2\in{\cal I}^2({\cal K},\Gamma)$ by
\[ V_0 = \sum_{n=1}^\infty |f_n\rangle\langle e_n|,\qquad
V_2 = \sum_{n=1}^\infty |e_{n+2}\rangle\langle e_n| \]
such that
\[ V=V_0 V_2^N, \qquad \varrho_V=\varrho_{V_0}
\varrho_{V_2}^N. \]
Since $V_0$ is unitary $P_0=V_0^*PV_0$ is again
a basis projection, and
\[ \pi_P\circ\varrho_V=\pi_{P_0}\circ\varrho_{V_2}^N. \]
Now we can use the foregoing results iteratively. Since
$\pi_{P_0}$ is a Fock representation
$\pi_{P_0}\circ\varrho_{V_2}$ decomposes into two
equivalent Fock representations, say
\[ \pi_{P_0}\circ\varrho_{V_2}=\pi_{P_1^{(1)}}
\oplus\pi_{P_2^{(1)}} \]
where, using Eq.~(\ref{basis}), in any case
$P_j^{(1)}-V_2^*P_0V_2$ is Hilbert
Schmidt class, $j=1,2$. In the next step, we find
a decomposition into four equivalent Fock
representations,
\[ \pi_{P_0}\circ\varrho_{V_2}^2=\pi_{P_1^{(1)}}\circ
\varrho_{V_2} \oplus \pi_{P_2^{(1)}}\circ\varrho_{V_2}
=\pi_{P_1^{(2)}} \oplus \pi_{P_2^{(2)}} \oplus
\pi_{P_3^{(2)}} \oplus \pi_{P_4^{(2)}} \]
where in any case $P_j^{(2)}-(V_2^*)^2P_0V_2^2$ is
Hilbert Schmidt class, $j=1,2,3,4$, and so on.
At the end one finds
\[ \pi_P\circ\varrho_V=\pi_{P_0}\circ\varrho_{V_2}^N
= \bigoplus_{j=1}^{2^N} \pi_{P_j^{(N)}} \]
where $P_j^{(N)}-(V_2^*)^NP_0V_2^N=P_j^{(N)}-V^*PV$ is
Hilbert Schmidt class, $j=1,2,\ldots,N$. Using Theorem
\ref{Hilbert Schmidt},
we obtain that all representations $\pi_{P_j^{(N)}}$
must be mutually equivalent. The proof is complete,
q.e.d.

Because of the decomposition of $\pi_P\circ\varrho_V$
into $2^N$ irreducibles, there must exist a set of
$2^N$ disjoint projections in ${\cal B}({\cal H}_P)$
(the projections onto invariant
subspaces), commuting with
$\pi_P\circ\varrho_V({\cal C}({\cal K},\Gamma))$,
which sum up to unity. To complete the picture,
we construct such a set. By setting
\[ g_{\pm j}=\frac{1}{\sqrt{2}}(f_j \pm if_{N+j}),
\qquad j=1,2,\ldots,N, \]
we find an ONB $\{g_j,j=\pm 1,\pm 2,\ldots,\pm N\}$ of
ker$V^*$ with the property $g_j=\Gamma g_{-j}$,
$j=1,2,\ldots N$. We define the mutually
commuting projections
\[ \Pi_j^\pm = \pi_P(B(g_{\pm j})^*B(g_{\pm j}))
\in \pi_P\circ\varrho_V({\cal C}({\cal K},\Gamma))'\]
such that $\Pi_j^++\Pi_j^-={\bf 1}$, $j=1,2,\ldots,N$.
Then we have $2^N$ projections of the form
\[ \Pi_{\epsilon_1,\epsilon_2,\ldots,\epsilon_N}=
\Pi_1^{\epsilon_1} \Pi_2^{\epsilon_2} \cdots
\Pi_N^{\epsilon_N},\qquad \epsilon_j=\pm,\qquad
j=1,2,\ldots,N \]
which have the desired properties.

\section{The odd case: $M_V=2N+1$}
Following Araki \cite{Ara1}, a projection
$F\in{\cal B}({\cal K})$ with the property
that $F\perp\Gamma F\Gamma$ and
\[ ({\bf 1}-F-\Gamma F\Gamma){\cal K}=
\mbox{span}\{f_0\}, \]
where $f_0\in{\cal K}$ is a normed,
$\Gamma$-invariant vector is called a partial
basis projection with $\Gamma$-codimension 1.
By $({\cal H}_F,\pi_F,|\Omega_F\rangle)$
we denote the Fock representation of
${\cal C}((F+\Gamma F\Gamma){\cal K},\Gamma)$
corresponding to $F$. (Note that $F$ is a
basis projection of
$((F+\Gamma F\Gamma){\cal K},\Gamma)$.)
It is proven in \cite{Ara1} that for such an
$F$ there exists an irreducible representation
$\pi_{(F,f_0)}$ of ${\cal C}({\cal K},\Gamma)$
on the Fock space ${\cal H}_F$, uniquely
determined by
\begin{equation}
\label{pi-F}
\pi_{(F,f_0)}(B(f))= \frac{1}{\sqrt{2}}
\langle f_0,f \rangle T_F(-1) + \pi_F (B(Ff+
\Gamma F \Gamma f)).
\end{equation}
Here $T_F(-1)$ denotes the unitary operator which
implements the automorphism $\alpha_{-1}(B(f))=-B(f)$
of ${\cal C}((F+\Gamma F\Gamma){\cal K},\Gamma)$ in
$\pi_F$. To the representation $\pi_{(F,f_0)}$, there
corresponds the non-quasifree state of
${\cal C}({\cal K},\Gamma)$,
\[ \omega_{(F,f_0)}(x)= \langle \Omega_F
|\pi_{(F,f_0)} (x) | \Omega_F \rangle ,
\qquad x\in {\cal C}({\cal K},\Gamma). \]
Araki proved the following
\begin{lemma}
\label{partial}
Let $F$ be a partial basis projection with
$\Gamma$-codimension 1 and define
$S\in Q({\cal K},\Gamma)$ by
\begin{equation}
\label{S(F)}
S=\frac{1}{2}({\bf 1}+F-\Gamma F \Gamma).
\end{equation}
Then the quasifree state $\omega_S$ of
${\cal C}({\cal K},\Gamma)$ decomposes according to
\begin{equation}
\omega_S=\frac{1}{2}(\omega_{(F,f_0)}+
\omega_{(F,-f_0)} ).
\end{equation}
Pure states $\omega_{(F,f_0)}$ and $\omega_{(F,-f_0)}$
give rise to inequivalent representations.
\end{lemma}
Now suppose a given Bogoliubov operator
$W\in{\cal I}^1({\cal K},\Gamma)$ and an arbitrary basis
projection $P\in{\cal B}({\cal K})$. We claim that
$S=W^*PW$ is always of the form (\ref{S(F)}) so that
Lemma \ref{partial} can be applied.
\begin{lemma}
\label{WPW}
Let $P$ be a basis projection of $({\cal K},\Gamma)$
and $W\in{\cal I}^1({\cal K},\Gamma)$. Then there
exists a partial basis projection $F$ with
$\Gamma$-codimension 1 such that
\begin{equation}
S=W^*PW=\frac{1}{2}({\bf 1}+F-\Gamma F\Gamma).
\end{equation}
By Lemma \ref{partial} the state
$\omega_S=\omega_P\circ\varrho_W$ splits into
two pure states. Furthermore, the representation
$\pi_P\circ\varrho_W$ decomposes into two
inequivalent representations.
\end{lemma}
{\it Proof.} Let $g_0$ be the $\Gamma$-invariant,
normed vector which spans ker$W^*$, and we
introduce $G_0=|g_0\rangle\langle g_0|$. We
then define
\[ E=4(S-S^2)=4(W^*PW-W^*PWW^*PW)=4W^*P
({\bf 1}-WW^*)PW=4WPG_0PW. \]
Since $g_0$ is $\Gamma$-invariant we find
\[ \langle g_0,Pg_0 \rangle = \langle \Gamma g_0,P
\Gamma g_0 \rangle = \langle ({\bf 1}-P)g_0,g_0
\rangle = 1-\langle g_0,Pg_0 \rangle \]
and therefore
\[ \langle g_0,P g_0 \rangle = \frac{1}{2},\qquad
G_0PG_0= \frac{1}{2} G_0. \]
This leads us to the fact that $E$ is a projection,
\[ E^2=16 W^*PG_0PWW^*PG_0PW = 16 W^*PG_0P
({\bf 1}-G_0)PG_0PW = 2E-E =E. \]
On the other hand we find $\Gamma E\Gamma=E$,
\[ \Gamma E\Gamma = 4 W^*\Gamma PG_0P\Gamma W
= 4W^*({\bf 1}-P)G_0({\bf 1}-P)W = 4W^*PG_0PW=E \]
since $W^*G_0=0=G_0W$ (remember that $G_0$ is the
projection onto ker$W^*$). Moreover, we compute
\[ ES=4W^*PG_0PWW^*PW=4W^*PG_0P({\bf 1}-G_0)PW
=E-\frac{1}{2}E=\frac{1}{2}E, \]
and also $SE=\frac{1}{2}E$. Now we define
\[ F=S-\frac{1}{2}E. \]
$F$ is a projection,
\[ F^2=S^2-\frac{1}{2}ES-\frac{1}{2}SE+\frac{1}{4}E^2
=(S-\frac{1}{4}E)-\frac{1}{4}E-\frac{1}{4}E+
\frac{1}{4}E=F, \]
and we have the relation
\[ {\bf 1}-F-\Gamma F \Gamma = {\bf 1} - (S-\frac{1}{2}E)
-({\bf 1}-S-\frac{1}{2}E) =E. \]
Moreover, we have $F\perp\Gamma F\Gamma$ since
\begin{eqnarray*}
F\Gamma F\Gamma &=& (S-\frac{1}{2}E)({\bf 1}-S-
\frac{1}{2}E)=S-S^2-\frac{1}{2}SE-\frac{1}{2}E
+\frac{1}{2}ES+\frac{1}{4}E \\
&=& \frac{1}{4}E-\frac{1}{4}E-\frac{1}{2}E+
\frac{1}{4}E+\frac{1}{4}E=0.
\end{eqnarray*}
For proving that $F$ is a partial basis projection with
$\Gamma$-codimension 1, it remains to be shown that $E$
is a rank-one-projection. We do that by computing that
its trace is one,
\begin{eqnarray*}
\mbox{tr}(E) &=& 4\,\mbox{tr}(W^*PG_0PW)= 4\,\mbox{tr}
(PG_0PWW^*) = 4\,\mbox{tr}(PG_0P({\bf 1}-G_0))\\
&=& 4\,\mbox{tr}(G_0P) - 2\,\mbox{tr}(PG_0)
= 2\,\mbox{tr}(PG_0) = 2 \sum_{n=0}^\infty
\langle g_n, P G_0 g_n \rangle = 2 \langle g_0,Pg_0
\rangle = 1
\end{eqnarray*}
where $\{g_n,n\in{\bf N}_0\}$ is any ONB of ${\cal K}$
containing $g_0$. Now the conditions for the
application of Lemma \ref{partial} are fulfilled, therefore
$\omega_P\circ\varrho_W$ decomposes into two pure
states, giving rise to inequivalent representations.
But since $\langle g_0,Pg_0 \rangle=\frac{1}{2}$
and the normed vector $g_0$ spannes the kernel of
$W^*$, we find ker$W^*\cap P{\cal K}=\{0\}$ and
therefore
$({\cal H}_P,\pi_P\circ\varrho_W,|\Omega_P\rangle)$
is a GNS representation of the state
$\omega_P\circ\varrho_W$ by Lemma \ref{Binnenhei}.
It follows that $\pi_P\circ\varrho_W$ decomposes into
two inequivalent representations, q.e.d.

Now we come to the general case that the corank of
a Bogoliubov operator $V$ is odd.
\begin{theorem}
\label{odd}
Let $\omega_P$ be a Fock state and let
$({\cal H}_P,\pi_P,|\Omega_P\rangle)$ be
the corresponding Fock representation
of the selfdual CAR algebra
${\cal C}({\cal K},\Gamma)$. Further let
$V\in{\cal I}({\cal K},\Gamma)$ be a
Bogoliubov operator with finite odd corank,
i.e. $M_V=2N+1$, $N\in{\bf N}_0$,
and let $\varrho_V$ be the
corresponding Bogoliubov endomorphism.
Then the representation $\pi_P\circ\varrho_V$
decomposes into $2^{N+1}$ irreducible
subrepresentations, namely we have
\begin{equation}
\label{decomodd}
\pi_P\circ\varrho_V= \left( \bigoplus_{j=1}^{2^N}
\pi_+^{(j)} \right) \oplus \left(
\bigoplus_{j=1}^{2^N} \pi_-^{(j)} \right).
\end{equation}
Representations $\pi_\pm^{(j)}$ are not Fock
representations. Moreover, for all $j,j'=1,2,
\ldots,2^N$, representations $\pi_+^{(j)}$ and
$\pi_+^{(j')}$ are unitarily equivalent, also
$\pi_-^{(j)}$ and $\pi_-^{(j')}$ are unitarily
equivalent but representations $\pi_+^{(j)}$ and
$\pi_-^{(j')}$ are inequivalent, and, since
irreducible, disjoint.
\end{theorem}
{\it Proof.} We choose again a $\Gamma$-invariant ONB
$\{e_n,n\in{\bf N}\}$ of ${\cal K}$, $e_n=\Gamma e_n$,
$n\in{\bf N}$. Furthermore, we choose a
$\Gamma$-invariant ONB $\{f_n,n=0,1,\ldots,2N\}$
of ker$V^*$, $f_n=\Gamma f_n$, $n=0,1,\ldots,2N$.
We define
\[ f_{2N+n}=V e_n,\qquad n=1,2,\ldots \]
Then $\{f_n,n\in{\bf N}_0\}$ is an ONB of ${\cal K}$, too,
and we can write
\[ V=\sum_{n=1}^\infty |f_{2N+n}\rangle\langle e_n|. \]
We introduce Bogoliubov operators
$W_1\in{\cal I}^1({\cal K},\Gamma)$ and
$W_2\in{\cal I}^2({\cal K},\Gamma)$,
\[ W_1 = \sum_{n=1}^\infty  |f_n\rangle\langle e_n|,
\qquad W_2 = \sum_{n=0}^\infty |f_{n+2}\rangle
\langle f_n|. \]
such that
\[ V=W_2^N W_1,\qquad \varrho_V = \varrho_{W_2}^N
\varrho_{W_1}. \]
By Theorem \ref{even} it follows that
\[ \pi_P\circ\varrho_V = \bigoplus_{j=1}^{2^N}
\pi_{P_j}\circ\varrho_{W_1} \]
where $\pi_{P_j}$, $j=1,2,\ldots,2^N$ are mutually
equivalent irreducible (Fock) representations. Now
we can use Lemma \ref{WPW}:
Each $\pi_{P_j}\circ\varrho_{W_1}$ decomposes into
two inequivalent irreducible (non-Fock)
representations,
\[ \pi_{P_j}\circ\varrho_{W_1} = \pi_+^{(j)} \oplus
\pi_-^{(j)}. \]
Since the representations $\pi_{P_j}\circ\varrho_{W_1}$
are mutually equivalent, we can choose the $\pm$-sign
such that $\pi_+^{(j)}$ and $\pi_+^{(j')}$ are
equivalent, also $\pi_-^{(j)}$ and $\pi_-^{(j')}$, and
that $\pi_+^{(j)}$ and $\pi_-^{(j')}$ are disjoint,
$j,j'=1,2,\ldots,2^N$, q.e.d.

We have seen that in the case
$V\in{\cal I}^{2N+1}({\cal K},\Gamma)$ the
representation $\pi_P\circ\varrho_V$ is a
direct sum of $2^{N+1}$ irreducible representations.
Thus there must exist a set of $2^{N+1}$
disjoint projections in ${\cal B}({\cal H}_P)$,
commuting with
$\pi_P\circ\varrho_V({\cal C}({\cal K},\Gamma))$,
which sum up to unity. By setting
\[ g_0=f_0,\qquad g_{\pm j}= \frac{1}{\sqrt{2}}
(f_j \pm i f_{N+j}), \qquad j=1,2,\ldots,N, \]
we find an ONB $\{g_j,j=0,\pm 1,\ldots,\pm N\}$
of ker$V^*$ with the property $g_j=\Gamma g_{-j}$,
$j=0,1,\ldots,N$. Let again
$T_P(-1)\in{\cal B}({\cal K})$ be a unitary
operator which implements the automorphism
$\alpha_{-1}$ of ${\cal C}({\cal K},\Gamma)$,
Eq.~(\ref{alpha}), in the Fock representation
$\pi_P$. Since $\alpha_{-1}^2=id$
the unitary $T_P(-1)$ can be choosen to be
selfadjoint, i.e. $T_P(-1)^2={\bf 1}$. We have
\[ \pi_P(B(g_0))T_P(-1)=-T_P(-1)\pi_P(B(g_0))\in
\pi_P\circ\varrho_V({\cal C}({\cal K},\Gamma))', \]
and by $B(g_0)^2=\frac{1}{2}{\bf 1}$ we find
that
\[ \Pi_0^\pm = \frac{1}{2} ({\bf 1} \pm \sqrt{2} i
\pi_P(B(g_0)) T_P(-1)) \]
are disjoint projections in
$\pi_P\circ\varrho_V({\cal C}({\cal K},\Gamma))'$.
Also the projections
\[ \Pi_j^\pm = \pi_P(B(g_{\pm j})^*B(g_{\pm j})),
\qquad j=1,2,\ldots,N \]
lie in the commutant of
$\pi_P\circ\varrho_V({\cal C}({\cal K},\Gamma))$,
one has $\Pi_j^++\Pi_j^-={\bf 1}$ for
$j=0,1,2,\ldots,N$, and all these projections
commute mutually. Now we are able to construct
$2^{N+1}$ projections
\[ \Pi_{\epsilon_0,\epsilon_1,\ldots,\epsilon_N}=
\Pi_0^{\epsilon_0} \Pi_1^{\epsilon_1} \cdots
\Pi_N^{\epsilon_N}, \qquad \epsilon_j=\pm,\qquad
j=0,1,\ldots,N \]
which have the desired properties.

Because the decomposition of the representation
$\pi_P\circ\varrho_V$ into irreducible
subrepresentations contains here two different
equivalence classes we simply conclude that
$\pi_P\circ\varrho_V$ cannot be unitarily
equivalent to a multiple of $\pi_P$.
\begin{corollary}
\label{nixquasi}
If $V\in{\cal I}({\cal K},\Gamma)$ is a
Bogoliubov operator with finite odd corank,
i.e. $M_V=2N+1$, $N\in{\bf N}_0$,
then representations $\pi_P$ and
$\pi_P\circ\varrho_V$ cannot be
quasiequivalent for any Fock representation
$\pi_P$ of ${\cal C}({\cal K},\Gamma)$.
\end{corollary}
We remark that this corollary agrees with
Binnenhei's recent results on isometrical
implementability of Bogoliubov endomorphisms
\cite{Binnenneu}.

\section{Restriction to the Even Subalgebra}
We now are interested in what happens when our
representations of ${\cal C}({\cal K},\Gamma)$
become restricted to the even subalgebra
${\cal C}({\cal K},\Gamma)^+$ which is the
algebra of fixpoints under the automorphism
$\alpha_{-1}$ of Eq.~(\ref{alpha}),
\[ {\cal C}({\cal K},\Gamma)^+ = \{ x\in
{\cal C}({\cal K},\Gamma)\,|\,\alpha_{-1}
(x)=x \}. \]
We begin with a lemma which is taken from
Araki's work \cite{Ara2}.
\begin{lemma}
\label{split}
Let $({\cal H}_P,\pi_P,|\Omega_P\rangle)$
be a Fock representation of
${\cal C}({\cal K},\Gamma)$. In the restriction
to the even subalgebra
${\cal C}({\cal K},\Gamma)^+$, the representation
$\pi_P$ splits into two irreducible
subrepresentations,
\begin{equation}
\pi_P|_{{\cal C}({\cal K},\Gamma)^+}=
\pi_P^+ \oplus \pi_P^-,
\end{equation}
Representations $\pi_P^+$ and $\pi_P^-$ are
disjoint. The commutant of
$\pi_P({\cal C}({\cal K},\Gamma)^+)$ is
generated by the $\alpha_{-1}$-implementing
$T_P(-1)$. The unitary $T_P(-1)$ can be
choosen to be selfadjoint, i.e.
$T_P(-1)^2={\bf 1}$.
\end{lemma}
Remember that for a Bogoliubov operator $V$ with
$M_V=2N$ a representation $\pi_P\circ\varrho_V$
splits into $2^N$ mutually
equivalent Fock representations. Therefore
$\pi_P\circ\varrho_V$, when restricted to
the even subalgebra, decomposes into $2^{N+1}$
irreducibles where one has two different
equivalence classes of $2^N$ mutually equivalent
subrepresentations each. Now suppose
$M_V=2N+1$. What happens with representations
$\pi_\pm^{(j)}$ of Theorem \ref{even},
Eq.~(\ref{decomodd}), in the restriction to
the even subalgebra? They are associated to
irreducible subrepresentations
$\pi_{(F,\pm f_0)}$, Eq.~(\ref{pi-F}), of
representations $\pi_S$ of
${\cal C}({\cal K},\Gamma)$, with corresponding
quasifree states $\omega_S$, $S$ of the form
(\ref{S(F)}). We will find that the situation
becomes completely inverted: An irreducible
Fock representation $\pi_P$ splits, when
restricted to ${\cal C}({\cal K},\Gamma)^+$,
into two inequivalent irreducibles. On the
other hand, such a representation $\pi_S$ splits
already as representation of
${\cal C}({\cal K},\Gamma)$ into
two inequivalent irreducible subrepresentatins,
however, these subrepresentations, when
restricted to ${\cal C}({\cal K},\Gamma)^+$,
remain irreducible but become equivalent.
\begin{lemma}
\label{equiodd}
The irreducible representation $\pi_{(F,f_0)}$,
Eq.~(\ref{pi-F}), of ${\cal C}({\cal K},\Gamma)$
remains irreducible when restricted to the even
subalgebra ${\cal C}({\cal K},\Gamma)^+$.
Moreover, in the restriction to
${\cal C}({\cal K},\Gamma)^+$, representations
$\pi_{(F,f_0)}$ and $\pi_{(F,-f_0)}$ become
equivalent.
\end{lemma}
{\it Proof.} For proving the irreducibilty of the
restricted representation $\pi_{(F,f_0)}$, we
assume an operator $A\in{\cal B}({\cal H}_F)$
which commutes with every $\pi_{(F,f_0)}(x)$,
$x\in{\cal C}({\cal K},\Gamma)^+$. Then we
show that it follows $A=\lambda{\bf 1}$,
$\lambda\in{\bf C}$, immediately. Since
$A\in\pi_{(F,f_0)}({\cal C}({\cal K},\Gamma)^+)'$
the operator $A$ commutes, in particular, with
all representors of the subalgebra
${\cal C}((F+\Gamma F\Gamma){\cal K},\Gamma)^+$,
i.e.
\[ [A,\pi_{(F,f_0)}(x)]=0,\qquad x\in{\cal C}
((F+\Gamma F\Gamma){\cal K},\Gamma)^+. \]
But for $x\in{\cal C}((F+\Gamma F\Gamma)
{\cal K},\Gamma)^+$ we have
$\pi_{(F,f_0)}(x)=\pi_F(x)$ since the first
summand vanishes in Eq.~(\ref{pi-F}). However,
since $\pi_F$ is a Fock representation of
${\cal C}((F+\Gamma F\Gamma){\cal K},\Gamma)$
we conclude that the commutant of $\pi_F(
{\cal C}((F+\Gamma F\Gamma){\cal K},\Gamma)^+)$
is generated by $T_F(-1)$ by Lemma \ref{split}.
This leads us to the ansatz
\[ A= \lambda {\bf 1} + \mu T_F(-1), \qquad
\lambda,\mu\in{\bf C}. \]
Now choose a non-zero
$f\in(F+\Gamma F\Gamma){\cal K}$.
We consider the representor of
$B(f_0)B(f)\in{\cal C}({\cal K},\Gamma)^+$,
\[ \pi_{(F,f_0)}(B(f_0)B(f)) = \pi_{(F,f_0)}(B(f_0))
\pi_{(F,f_0)}(B(f)) = \frac{1}{\sqrt{2}} T_F(-1)
\pi_F(B(f)). \]
We compute
\begin{eqnarray*}
[ A,\pi_{(F,f_0)}(B(f_0)B(f)) ] &=&
[ \mu T_F(-1),\frac{1}{\sqrt{2}}T_F(-1)\pi_F(B(f)) ]\\
&=& \frac{\mu}{\sqrt{2}} (\pi_F(B(f))-T_F(-1)
\pi_F(B(f)) T_F(-1)) \\
&=& \sqrt{2} \mu \pi_F(B(f)).
\end{eqnarray*}
But since
$A\in\pi_{(F,f_0)}({\cal C}({\cal K},\Gamma)^+)'$
this commutator has to vanish. This implies
$\mu=0$ and therefore $A=\lambda{\bf 1}$. It
remains to be shown that $\pi_{(F,f_0)}$ and
$\pi_{(F,-f_0)}$, when restricted to
${\cal C}({\cal K},\Gamma)^+$,
become equivalent. Now choose arbitrary
$f_1,f_2\in{\cal K}$. By Eq.~(\ref{pi-F}) we
compute
\begin{eqnarray*}
\lefteqn{\pi_{(F,\pm f_0)}(B(f_1)B(f_2))=} \\
&=& \frac{1}{2}
\langle f_0,f_1 \rangle \langle f_0,f_2 \rangle
\pm \frac{1}{\sqrt{2}} \langle f_0,f_1 \rangle
T_F(-1) \pi_F(B(Ff_2+\Gamma F\Gamma f_2)) \\
&& \pm \frac{1}{\sqrt{2}} \langle f_0,f_2
\rangle \pi_F(B(Ff_1+\Gamma F\Gamma f_1)) T_F(-1)
+ \pi_F(B(Ff_1+\Gamma F\Gamma f_1))
\pi_F(B(Ff_2+\Gamma F\Gamma f_2)).
\end{eqnarray*}
Since $T_F(-1) \pi_F(B(Ff_j+\Gamma F\Gamma f_j))
=- \pi_F(B(Ff_j+\Gamma F\Gamma f_j))T_F(-1)$,
$j=1,2$, one sees easily
\[ \pi_{(F,f_0)}(B(f_1)B(f_2))=T_F(-1)
\pi_{(F,-f_0)}(B(f_1)B(f_2))T_F(-1). \]
Now ${\cal C}({\cal K},\Gamma)^+$ is generated
by such elements $B(f_1)B(f_2)$.
Therefore representations $\pi_{(F,f_0)}$
and $\pi_{(F,-f_0)}$ of
${\cal C}({\cal K},\Gamma)^+$ are unitarily
equivalent; the equivalence is realized
by $T_F(-1)$, q.e.d.

Now we can apply these results to representations
$\pi_P\circ\varrho_V$ where $V$ is a Bogoliubov
operator with finite corank. By applying
Lemmata \ref{split} and \ref{equiodd} to the
subrepresentations of $\pi_P\circ\varrho_V$
according to Theorems \ref{even} and \ref{odd}
we conclude
\begin{theorem}
\label{evenodd}
Let $\omega_P$ be a Fock state and let
$({\cal H}_P,\pi_P,|\Omega_P\rangle)$ be
the corresponding Fock representation
of the selfdual CAR algebra
${\cal C}({\cal K},\Gamma)$. Further let
$V\in{\cal I}({\cal K},\Gamma)$ be a
Bogoliubov operator with finite corank,
and let $\varrho_V$ be the corresponding
Bogoliubov endomorphism. If the corank
of $V$ is an even number, say $M_V=2N$,
$N\in{\bf N}_0$, then the representation
$\pi_P\circ\varrho_V$, when restricted
to the even subalgebra
${\cal C}({\cal K},\Gamma)^+$,
decomposes into $2^{N+1}$ irreducible
subrepresentations, namely we have
\begin{equation}
\pi_P\circ
\varrho_V|_{{\cal C}({\cal K},\Gamma)^+}
= \left( \bigoplus_{j=1}^{2^N}
\pi_{P_j}^+ \right) \oplus \left(
\bigoplus_{j=1}^{2^N} \pi_{P_j}^- \right).
\end{equation}
For all $j,j'=1,2,\ldots,2^N$, representations
$\pi_{P_j}^+$ and $\pi_{P_{j'}}^+$ are unitarily
equivalent, also $\pi_{P_j}^-$ and $\pi_{P_{j'}}^-$
are unitarily equivalent but representations
$\pi_{P_j}^+$ and $\pi_{P_{j'}}^-$ are disjoint.
On the other hand, if the corank of $V$ is an
odd number, say $M_V=2N+1$, $N\in{\bf N}_0$,
then the representation $\pi\circ\varrho_V$,
when restricted to ${\cal C}({\cal K},\Gamma)^+$,
decomposes into $2^{N+1}$ mutually equivalent
irreducible subrepresentations.
\end{theorem}

\section{Concluding Remarks}
We have learned into how many irreducible
subrepresentations a representation
$\pi_P\circ\varrho_V$ of CAR splits in the case
that the corank of the Bogoliubov operator $V$
is finite. Moreover, we know something about
the equivalence classes of those
subrepresentations. These results were gained
by a suitable product decomposition of the
Bogoliubov operator. Because one has to take
into account a lot of distinctions, concerning
the vanishing or non-vanishing of scalar
products $\lambda_\pm$ in each step, we did not
give explicit formulae for those subrepresentations
for the case that $V\in{\cal I}^n({\cal K},\Gamma)$,
$n>2$. However, suppose such a Bogoliubov operator
given explicitely, it can be seen from our proofs
how one has to construct the subrepresentations
(i.e. generating vectors (\ref{globvec}),(\ref{locvec})
and basis projections, Lemmata \ref{Binnenhei},
\ref{Bockenhauer}) step by step. Therefore one
always obtains an explicit
construction. (We remark that in the case
of equivalent subrepresentations, the decomposition
into invariant subspaces is not unique.)
Moreover, we have seen what happens with those
representations when they become restricted to
the even subalgebra. But what happens when
the corank of the Bogoliubov operator $V$ becomes
infinite? Our product decomposition of $V$ then
becomes infinite; one cannot receive explicit
formulae in this way, moreover, we cannot
say anything about the equivalence classes
of the subrepresentations of $\pi_P\circ\varrho_V$.
However, because an ONB of ker$V^*$ then becomes
infinite it is possible to construct an unbounded
number of disjoint projections in the commutant of
$\pi_P\circ\varrho_V({\cal C}({\cal K},\Gamma))$.
Clearly, the representation $\pi_P\circ\varrho_V$
possesses a decomposition into infinitely many
irreducibles.


\begin{thebibliography}{99}
\bibitem{Ara1} {\sc H. Araki:} {\sl On Quasifree States
of CAR and Bogoliubov Automorphisms.} Publ.~RIMS Kyoto
Univ.~Vol.~{\bf 6} (1970/71)
\bibitem{Ara2} {\sc H. Araki:} {\sl Bogoliubov Automorphisms
and Fock Representations of the Canonical Anticommutation
Relations.} In: Operator Algebras and Mathematical
Physics, Am.~Math.~Soc.~Vol.~{\bf 62} (1987)
\bibitem{PS} {\sc R.T. Powers, E. St{\o}rmer:} {\sl Free
States of the Canonical Anticommutation Relations.}
Commun.~Math.~Phys.~{\bf 16} (1970)
\bibitem{Rideau} {\sc G. Rideau:} {\sl On Some
Representations of the Anticommutation Relations.}
Commun.~Math. Phys.~{\bf 9} (1968)
\bibitem{Bin} {\sc C. Binnenhei:}
{\sl Bogoliubov-Transformationen und lokalisierte
Morphismen f{\"u}r freie Dirac-Felder.}
Diplomarbeit, Bonn (1993)
\bibitem{Binnenneu} {\sc C. Binnenhei:} {\sl Implementation
of Endomorphisms of the CAR algebra.} In preparation
\bibitem{ich} {\sc J. B{\"o}ckenhauer:} {\sl Localized
Endomorphisms of the Chiral Ising Model.} DESY-preprint
No. 94-116 (1994)
\end{thebibliography}
\end{document}